\documentclass[reprint,prb,aps,amsmath,amssymb,superscriptaddress,nofootinbib]{revtex4-2}
\usepackage[colorlinks, citecolor=blue, urlcolor=blue, linkcolor=red,breaklinks]{hyperref}
\usepackage{xcolor}
\usepackage{graphicx}
\usepackage{dcolumn}
\usepackage{bm}
\renewcommand{\vec}[1]{\boldsymbol{#1}}
\usepackage[normalem]{ulem}

\begin{document}

\preprint{APS/123-QED}


\title{Electronic Structure and Kohn-Luttinger Superconductivity of Heavily-Doped Single-Layer Graphene}

\author{Sa\'ul A. Herrera}
\affiliation{Depto. de Sistemas Complejos, Instituto de F\'isica, UNAM, Ciudad Universitaria, 04510 Ciudad de M\'exico, Mexico.}
\author{Guillermo Parra-Mart\'inez}
\affiliation{IMDEA Nanoscience, C/ Faraday 9, 28049 Madrid, Spain}
\author{Philipp Rosenzweig}
\affiliation{Max-Planck-Institut f\"ur Festk\"orperforschung, Heisenbergstr.\ 1, 70569 Stuttgart, Germany}
\affiliation{Physikalisches Institut, Universit\"at Stuttgart, Pfaffenwaldring 57, 70569 Stuttgart, Germany}
\author{Bharti Matta}
\affiliation{Max-Planck-Institut f\"ur Festk\"orperforschung, Heisenbergstr.\ 1, 70569 Stuttgart, Germany}
\author{Craig M.\ Polley}
\affiliation{MAX IV Laboratory, Lund University, Fotongatan 2, 22484 Lund, Sweden}
\author{Kathrin K\"uster}
\author{Ulrich Starke}
\affiliation{Max-Planck-Institut f\"ur Festk\"orperforschung, Heisenbergstr.\ 1, 70569 Stuttgart, Germany}
\author{Francisco Guinea}
\affiliation{IMDEA Nanoscience, C/ Faraday 9, 28049 Madrid, Spain}
\affiliation{Donostia International Physics Center, Paseo Manuel de Lardiz\'{a}bal 4, 20018 San Sebastián, Spain}
\author{Jos\'e \'Angel Silva-Guill\'en}
\email{joseangel.silva@imdea.org}
\affiliation{IMDEA Nanoscience, C/ Faraday 9, 28049 Madrid, Spain}
\author{Gerardo G. Naumis}
\affiliation{Depto. de Sistemas Complejos, Instituto de F\'isica, UNAM, Ciudad Universitaria, 04510 Ciudad de M\'exico, Mexico.}
\author{Pierre A. Pantale\'on}
\email{pierre.pantaleon@imdea.org}
\affiliation{IMDEA Nanoscience, C/ Faraday 9, 28049 Madrid, Spain}
 
\date{\today} 

\begin{abstract}

The existence of superconductivity (SC) in graphene appears to be established in both twisted and non-twisted multilayers. However, whether their building block, single-layer graphene (SLG), can also host SC remains an open question.\ Earlier theoretical works predicted that SLG could become a chiral $d$-wave superconductor driven by electronic interactions when doped to its van Hove singularity, but questions such as whether the $d$-wave SC survives the strong band renormalizations seen in experiments, its robustness against the source of doping, or if it will occur at any reasonable critical temperature ($T_c$) have remained difficult to answer, in part due to uncertainties in model parameters.\ In this study, we adopt a random-phase approximation framework based on a Kohn-Luttinger-like mechanism to investigate SC in heavily-doped SLG. We predict that robust $d+id$ topological SC could arise in SLG doped by Tb, with a $T_c$ up to 600 mK. We also investigate the possibility of realizing $d$-wave SC by employing other dopants, such as Li or Cs.\ The structural models have been derived from angle-resolved photoemission spectroscopy measurements on Tb-doped graphene and first-principles calculations for Cs and Li doping. We find that dopants that change the lattice symmetry of SLG are detrimental to the $d$-wave state.\ The stability of the $d$-wave SC predicted here in Tb-doped SLG could provide a valuable insight for guiding future experimental efforts aimed at exploring topological superconductivity in monolayer graphene.

\end{abstract}

\maketitle


\section{Introduction}

Since the discovery of superconductivity (SC) in twisted bilayer graphene (TBG) ~\cite{Cao2018Unconventional,Yankowitz2019Tuning,Lu2019Superconductors,Stepanov2020Untying,Oh2021evidence}, twisted trilayer graphene (TTG)~\cite{Park2021Tunable, Hao2021Electric,Kim2022evidence,Liu2022Isospin}, other twisted multilayers~\cite{Park2022Robust, Zhang2022Promotion}, non-twisted Bernal bilayer graphene (BBG)~\cite{Zhou2022Isospin,Zhang2023Enhanced,holleis2023Ising} and rhombohedral trilayer graphene (RTG)~\cite{Zhou2021Superconductivity}, substantial research has been dedicated to understand their different phases and electronic properties. In twisted systems, the fact that the SC state originates from narrow bands with large electronic interactions suggests an unconventional electronic mechanism~\cite{Gonzlez2019, Roy2019, Goodwin2019, Lewandowski2021, Sharma2020, Samajdar2020SC, Cea21Coulomb,Pahlevanzadeh2021DMFT, Crepel2022Unconventional,Gonzlez2023TTG}. 

Superconductivity has also been observed in graphite intercalation compounds (GICs)~\cite{Dresselhaus2002Intercalation,Smith2015Superconductivity,Takada2016Theory} and fullerene crystals doped with alkaline ions~\cite{Rosseinsky1991Superconductivity,Kelty1991Superconductivity,Chakravarty1991Superconductivity,Hebard1991Superconductivity,Hebard1992Superconductivity}.\ In these systems, there is no clear evidence for unconventional superconductivity, but the effect of electron-electron interactions has been considered in doped fullerenes~\cite{Capone2002Strongly,Nomura2015Unified,Calandra2005Theoretical,Wang1991Fermi,Cantaluppi2018Pressure,Haddon1986Electronic,Erwin1991Theoretical}.\ Although the mechanism leading to SC in twisted and non-twisted graphene heterostructures is still under debate, some of their superconducting phases exhibit signatures of unconventional pairing, which are not typically associated with phonon-driven SC. 

The growing list of superconducting graphene multilayers naturally leads to the question of whether their building block, i.e., graphene monolayer itself, can also host similar superconducting phases.\ Close to charge neutrality the graphene bands are highly dispersive, are well described in a single particle picture and form the well-known Dirac cones. However, at high doping levels, close to its van Hove singularity (VHS), many-body instabilities might be favored due to the high density of states (DOS) and the giant hole pocket in the Fermi surface (FS)~\cite{Black2007Resonating,Honerkamp2008Density,Gonzalez2008Kohn,Makogon2011Spin,Nandkishore2012Chiral,Kiesel2012Competing,Maiti2013Superconductivity,Kagan2014,BlackSchaffer2014Chiral}. Earlier theoretical works predicted single-layer graphene (SLG) could become an unconventional, chiral $d$-wave superconductor driven by electron interactions by raising its Fermi level ($E_F$) to the VHS~\cite{Black2007Resonating,Honerkamp2008Density,Gonzalez2008Kohn,McChesney2010Extended,Wang2012Functional,Nandkishore2012Chiral,Kiesel2012Competing,BlackSchaffer2014Chiral}.\ However, most predictions have been of qualitative character, and questions regarding the robustness of the $d$-wave SC against the source of doping, or whether it will occur at any reasonable critical temperature ($T_c$) have remained difficult to address~\cite{BlackSchaffer2014Chiral}.\ Due to a sensitive dependence between $T_c$ and coupling parameters, approximations in the theoretical framework and the use of simplified models to describe the system make it difficult to reliably estimate the $T_c$, as evidenced by past predictions ranging from a few, to hundreds of Kelvin~\cite{Gonzalez2008Kohn, Pathak2010Possible, Nandkishore2012Chiral,BlackSchaffer2014Chiral}.

\begin{figure}
\includegraphics[width=0.48\textwidth]{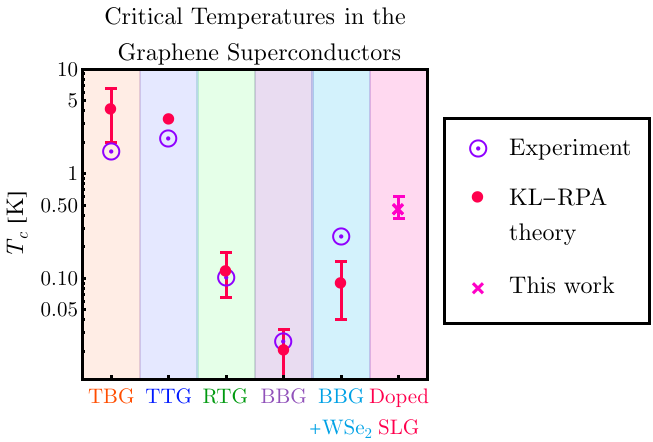}
\caption{Critical temperatures for graphene-based superconductors obtained from the calculation framework based on the Kohn-Luttinger mechanism used here and in our previous works \cite{Cea2021Coulomb,Phong2021Band,Cea2023Superconductivity,Jimeno2023Superconductivity,Li2023Charge,Long2024Evolution}. Experimental reports available so far for the $T_c$ \cite{Cao2018Unconventional,Zhou2021Superconductivity,Zhou2022Isospin,Park2021Tunable,Zhang2023Enhanced} are also shown. Notice the logarithmic scale. Good agreement between theory and experiment is seen across three orders of magnitude. Our prediction for the $T_c$ in doped SLG is also shown. Intervals indicate the variation in the computed $T_c$ obtained within different approximations and models.}
\label{Fig: Comparisons} 
\end{figure}

\begin{figure*}
\includegraphics[width=1\textwidth]{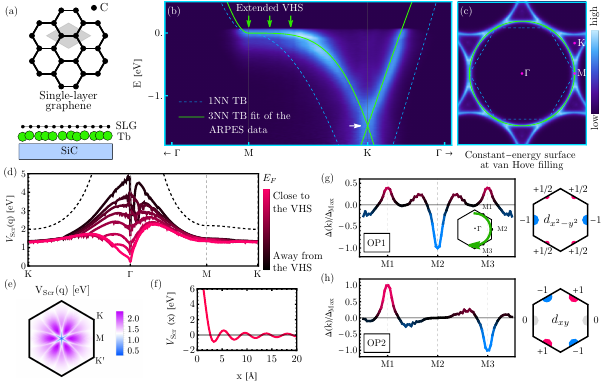}
\caption{
Tb intercalated SLG on SiC: (a) Model sketch of SLG doped by Tb intercalation with the $1\times 1$ lattice symmetry of SLG preserved (top), with Tb atoms intercalated as a monoatomic interface layer between (decoupled) SLG and the SiC substrate (bottom).\ (b) Renormalized $\pi^*$ band dispersion upon heavy doping by Tb intercalation measured by ARPES, featuring an extended VHS (green arrows), $E_F$ slightly above the VHS (by ${\approx}$ $0.07$ eV) and a reduced $\pi^*$ band width. The Dirac point (white arrow) lays at about $-1.55$ eV.
The renormalized dispersion is well reproduced by the overlaid TB model for SLG including up to third-nearest neighbors (3NN, solid green lines) while a 1NN TB model (dashed blue lines) does not capture the experimental band structure.\
(c) Rounded hole pocket in the constant energy surface at the VHS, overlaid with the 3NN TB fit (green line) and the 1NN TB fit (dashed lines), with some high-symmetry points of the $1\times1$ BZ indicated.\ (d-f) Calculated screened Coulomb interaction: (d) As graphene is doped close to the VHS, the electronic potential gets progressively screened close to $\Gamma$.\ Dashed line indicates the bare Coulomb interaction.\ (e) $V_{Scr}(q)$ in the full BZ.\ (f) In real space, $V_{Scr}$ reveals an effective attractive interaction. In panels (g) and (h) we display the gap variations throughout the BZ in the $d$-wave pairing channels $d_{x^2-y^2}$ and $d_{xy}$, respectively.\ The obtained symmetry of the OP is robust upon the strong band renormalizations induced by doping.}
\label{Fig: TerbiumSLG}
\end{figure*}

On the experimental side, important progress has been made in achieving the ideal doping levels for the superconducting state to arise. In a previous work \cite{McChesney2010Extended}, heavy doping of SLG was accomplished by employing a combination of intercalation and adsorption of calcium and potassium. Yet, the unambiguous demonstration of doping beyond the $\pi^*$ VHS in SLG was not realized until more recently by employing ytterbium intercalation and potassium adsorption \cite{Rosenzweig2020Overdoping}, reaching a previously inaccessible regime in the phase diagram of SLG, where exotic many-body states, such as $d$-wave SC, might emerge.

Importantly, experiments have shown that heavily doping SLG can significantly modify its electronic bands, in a way that is dopant-dependent \cite{McChesney2010Extended,Profeta2012Phonon,Ludbrook2015Evidence,Link2019Introducing,Rosenzweig2020Overdoping,Ehlen2020Origin,Bao2022Coexistence}. ARPES measurements have shown  that the most prevalent effect of a high electronic density in graphene is a strong renormalization of the $\pi$ bands, extending and flattening them close to the VHS, and rounding the FS~\cite{McChesney2010Extended, Link2019Introducing,Rosenzweig2020Overdoping}. This leads to an extended VHS (eVHS), also called a higher-order VHS, which is expected to have an impact on the competition between SC and other phases \cite{Classen2020Competing}. Similar features have been observed in cuprates~\cite{Gofron1994Observation} and other highly correlated materials~\cite{Lu1996Fermi}, and a higher-order VHS has been discussed more recently for TBG~\cite{Yuan2019Magic}.
A further complication arises in some cases where the dopants order periodically, forming dispersive bands that hybridize with the carbon states close to $E_F$, inducing a periodic potential that changes the lattice symmetry of SLG. 
This drastically modifies the electronic structure of pristine SLG. Such is the case for dopants as Li \cite{Ludbrook2015Evidence,Ichinokura2022Van,Bao2022Coexistence} and Cs \cite{Ehlen2020Origin}, where the electronic spectrum resembles less that of SLG and more that of GICs, some of which are conventional phonon-driven superconductors \cite{Dresselhaus2002Intercalation,Profeta2012Phonon}.

In this work, we derive realistic models for heavily-doped SLG and estimate the $T_c$ and order parameter (OP) of the superconducting state that might arise at fillings close to the VHS. We adopt a random-phase approximation (RPA) framework based on a Kohn-Luttinger-like (KL) mechanism which considers direct electronic interactions \cite{Cea2021Coulomb,Phong2021Band,Cea2023Superconductivity,Jimeno2023Superconductivity,Li2023Charge,Long2024Evolution}.\ This framework has recently been shown to lead to estimations of $T_c$ for non-twisted graphene multilayers that are in good agreement with experiments~\cite{Cea2023Superconductivity,Li2023Charge} (see also~\cite{Pantaleon2023Superconductivity} and references therein).\ Moreover, it has also been shown that it yields 
critical temperatures in agreement with experiments in TBG~\cite{Cao2018,Yankowitz2019,Lu2019_SC_TBG,Cea21Coulomb}, TTG~\cite{Park2021_SC_TTG, Hao2021_SC_TTG,Phong2021Band}, RTG~\cite{Zhou2021SuperRTG,Cea2022Superconductivity}, BBG~\cite{Zhou2022Isospin,zhang2022Spin,holleis2023Ising,Jimeno2023Superconductivity}, twisted double bilayer graphene~\cite{Su2023Superconductivity,Long2024Evolution} and, most recently, has allowed us to predict SC in helical TTG~\cite{Long2024Evolution}.\ A comparison between predictions made with the KL-RPA framework and experiments is shown in Fig.~\ref{Fig: Comparisons}. More details are given in  Methods sec.~\ref{sec: MethodSC}.

To faithfully represent the band structure of heavily doped graphene, we take ARPES measurements on Terbium-intercalated SLG, which leads to doping beyond the VHS of the monolayer. We also employ density functional theory (DFT) to calculate the electronic structure of alkali-doped SLG, which have been shown experimentally to further change the lattice symmetry of SLG by inducing $2\times 2$ \cite{Ehlen2020Origin} or $\sqrt{3}\times\sqrt{3}$~\cite{Ludbrook2015Evidence,Bao2022Coexistence} superlattice structures. We employ different tight-binding (TB) models to correctly describe each of the electronic dispersions, considering experimentally observed dopant-dependent features such as band flattening, dopant-carbon hybridization, and Brillouin zone (BZ) folding. We then determine the expected $T_c$ and OP of the SC phase employing the KL-RPA framework.\ Given that different dopants modify the band structure of SLG in distinct ways, we find that electron-driven SC can potentially arise only for certain types of chemical doping.\ Interestingly, our results suggest that when SLG is doped to the VHS, it becomes a chiral $d$-wave topological superconductor with a $T_c$ ranging from $\sim$ 370 to 600 mK, as long as there is no BZ folding induced by the dopants. Based on this criterion, dopants such as Tb, Yb~\cite{Rosenzweig2020Overdoping} or Gd~\cite{Link2019Introducing} might be the best candidates for achieving electron-driven topological superconductivity in SLG.

\section{Results} \label{sec: results}
\subsection{Tb-doped SLG} \label{Sec: results-tbslg}
First, as an example for graphene without a superlattice, we consider doping of SLG on SiC via Tb intercalation (Tb-SLG, see Methods sec. \ref{sec: graphene-exp}). This system's advantage is, that the VHS scenario is already reached and even surpassed by the doping through the interlayer alone, so that no additional charge from a top adsorbate is necessary. The atomic arrangement is such that the Tb does not hybridize with graphene and the primitive unit cell of graphene [see Fig.~\ref{Fig: TerbiumSLG}(a)] is preserved without any additional long range order. As noted, there is a substantial charge transfer from Tb onto SLG, pushing the $E_F$ from the Dirac point [white arrow in Fig.~\ref{Fig: TerbiumSLG}(b)] to $\approx0.07$ eV above the VHS (green arrows).\ This corresponds to an electron density of about $5\times10^{14}$~cm$^{-2}$ as obtained from the area enclosed by the giant hole pocket around $\Gamma$.\ Instead of a rigid shift of $E_F$, however, our ARPES measurements show a strong renormalization of the graphene bands, which are flattened and exhibit a higher-order, or extended, VHS \cite{Yuan2019Magic,Classen2020Competing}. As shown in Fig.~\ref{Fig: TerbiumSLG}(c), right at van Hove filling (i.e., $0.07$ eV below $E_F$) this results in a rounded energy surface contour touching the KMK$^\prime$ BZ edge. Such a renormalization resembles that of high-$T_c$ superconductors \cite{Gofron1994Observation,Lu1996Fermi,Irkhin2002Robustness}, and seems to be an intrinsic effect of VHS-doped SLG, as has also been obtained in experiments employing Ca~\cite{McChesney2010Extended}, Gd~\cite{Link2019Introducing} and Yb~\cite{Rosenzweig2019Tuning}.

The origin of the renormalization of $\pi^*$ states has been attributed to electron correlations~\cite{Link2019Introducing,Yudin2014Fermi}, and also to
the combination of e-e and e-ph interactions~\cite{McChesney2010Extended}. However, regardless of its detailed many-body nature, and in
order to consider its impact on the SC phase, we follow Refs.~\cite{McChesney2010Extended,Classen2020Competing} to take into account this renormalization
by fitting a third nearest-neighbor (3NN) effective single-particle TB Hamiltonian (see Methods sec. \ref{sec: TBFit}).
As indicated by the solid green curves in Fig. \ref{Fig: TerbiumSLG}(b) and (c) the TB model captures the band flattening and the
rounded constant-energy contour as measured by ARPES. Moreover, it displays a proper extended VHS in contrast to
the ordinary case of nearest-neighbor (1NN) TB [dashed curves in Fig. 2(b) and (c)] and yields the correct energy position of the Dirac point at -1.55 eV.

Electron doping moves the Fermi level close to the extended VHS and thus enhances the role of electronic interactions as visible in Fig.~\ref{Fig: TerbiumSLG}(b) and (c). We have employed the KL-RPA framework using this 3NN TB model.\ Results for the screened interaction and the superconducting OP of Tb-SLG are depicted in Fig.~\ref{Fig: TerbiumSLG}(d) to Fig.~\ref{Fig: TerbiumSLG}(h).\ The screened Coulomb potential (see Methods in sec.~\ref{sec: MethodSC}) as a function of $E_F$ is shown in Fig.~\ref{Fig: TerbiumSLG}(d-e), where a strong screening is obtained as the Fermi energy approaches the VHS.\ Fig.~\ref{Fig: TerbiumSLG}(e) displays its behavior within the BZ, where a clear minimum is obtained near $\Gamma$.\ The vanishing of $V_{Scr}(q)$ at the $\Gamma$ center indicates that the susceptibility diverges as $q \rightarrow 0$, leading to a locally attractive interaction in real space, as shown in Fig.~\ref{Fig: TerbiumSLG}(f).\ This strong screening, favorable for SC, results in a first minimum at $x \approx 3$~\AA~with an attractive strength of $V_{Scr} \approx -1.0$~eV shown in Fig.~\ref{Fig: TerbiumSLG}(f), resulting in a $T_c\approx 375$ mK positioning it among one of the strongest non-twisted graphene superconductors, with a $T_c$ more than twice as high as for RTG ($T^{RTG}_c\approx 150$ mK)~\cite{Zhou2021SuperRTG}.

We have found that, in contrast with other non-twisted graphene systems~\cite{Pantaleon2023Superconductivity}, the superconducting OP for Tb-SLG is doubly degenerated into $d_{x^2-y^2}$ and $d_{xy}$ pairing channels, each with a superconducting gap varying around the FS as shown in Fig.~\ref{Fig: TerbiumSLG}(g-h).\ This SC state was predicted in previous works~\cite{Black2007Resonating,Gonzalez2008Kohn,Nandkishore2012Chiral}.\ We corroborate its robustness against the band renormalizations seen in our ARPES experiment.\ It can be shown, via a free-energy analysis~\cite{Nandkishore2012Chiral}, that if a system possesses both $d$-wave pairing channels, the most favorable state involves a complex combination of both orders, giving rise to an exotic spin singlet $d+id$ topological superconductor~\cite{Fu2008Superconducting,Yanase2022Topological,Nandkishore2012Chiral,Vafek2012Carbon}. Since the $d+ id$ pairing produces a fully gapped state, it is energetically favorable \cite{BlackSchaffer2014Chiral}. Importantly, this state has a non-trivial topology \cite{BlackSchaffer2014Chiral} and it is expected to host Majorana modes under some additional modifications~\cite{Black2012Edge,Li2023High}.

\begin{figure*}
\includegraphics[width=1\textwidth]{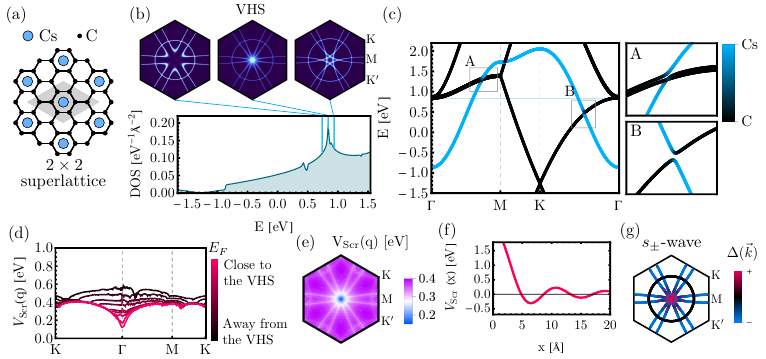}
\caption{Model for Cs-doped SLG and calculated screened electron-electron interaction. (a) $2\times2$ graphene-cesium superlattice. (b) DOS and Fermi surfaces at various fillings across the VHS, the latter corresponding to the filling with the highest DOS. High-symmetry points in all panels correspond to the first BZ of the $2\times2$ superlattice. (c) Band structure from a tight-binding model considering the $\pi^*$ orbitals of graphene and the $s$ orbital of cesium fitted from DFT calculations. The model captures the zone folding and hybridization features observed in ARPES measurements in Ref. \cite{Ehlen2020Origin}. The A and B insets show the hybridization between carbon and cesium bands (indicated by color) at the areas enclosed by the gray rectangles in the main panel.  Experiments in Ref. \cite{Ehlen2020Origin} show fillings close to the blue horizontal line. (d-f) Screened electron-electron interaction. (d) Coulomb interaction along a path in the first BZ. Interaction gets progressively screened as filling gets closer to the VHS. (e) Screened interaction in the full first BZ at the VHS. (f) Screened interaction in real space. (g) Leading OP at $T=10$ mK.}
\label{Fig: CesiumSLG}
\end{figure*}

In order to test the robustness of our results against variations in the parameters of the TB model, we carried out the same calculation for different sets of parameters.\ We find that if the hopping integrals are tuned such that the band becomes flatter while maintaining the extended VHS \cite{Classen2020Competing}, the $T_c$ is increased.\ However, such alternative sets of parameters fit the ARPES data poorly away from $E_F$ (see Methods sec. \ref{sec: TBFit}).\
We find an increase of the $T_c$ by a factor of $\approx 1.8$ at most ($T_c\approx600$ mK), while the OP is essentially unaltered.\ Importantly, this shows that the order of magnitude of $T_c$ and the symmetry of the OP remain robust against variations of the parameters of the model, suggesting that Tb-doped SLG might be a robust topological superconductor~\cite{Sato2017Topological}.

Regarding the effect that the e-ph coupling might have on the superconducting state, we note that ARPES experiments \cite{Park2008Van,McChesney2010Extended,Rosenzweig2020Overdoping,Jugovac2022Clarifying} consistently show a kink that arises $\approx$0.1-0.2 eV below the VHS due to mass renormalization from e-ph coupling. However, considering the estimated e-ph coupling constant \cite{Rosenzweig2020Overdoping,Bao2022Coexistence}, we do not expect a phonon-driven SC state to override the electron-driven $d$-wave state (see Methods sec. \ref{Subsec: phonons}).


\subsection{Doped Graphene Superlattices}

Employing different species of dopants might induce more significant changes into the electronic spectrum of SLG.\ Recent experimental reports have achieved high doping in monolayer and few-layer graphene employing Li \cite{Bao2022Coexistence,Jugovac2022Clarifying,Ichinokura2022Van} and Cs \cite{Ehlen2020Origin} intercalation and adsorption. Because those systems also exhibit a flat electronic band close to the $E_F$, their potential to host many-body instabilities, including SC, was suggested.\ However, the electronic structures of Li-doped and Cs-doped graphene differ significantly from the one we report for Tb-doped SLG.\ In Li and Cs doped graphene dopants order periodically, which tends to modify the electronic spectrum of graphene in two significant ways.\ First, the ordered dopants might induce a periodic potential that changes the lattice symmetry of graphene and folds its bands.\ Second, the ordered dopants might induce a free-electron-like interlayer band that hybridizes with the $\pi^*$ carbon band close to the $E_F$.\ In what follows we investigate whether the $d$-wave SC state can be expected to survive such dopant-induced features.

First we note that, in contrast to Tb-doped SLG, the electronic structure of Li and Cs doped graphene resembles that of GICs, where dopants intercalate in periodic arrangements between graphite layers \cite{Dresselhaus2002Intercalation}.\ In the GICs, the presence of an interlayer band crossing the Fermi level has been linked to the existence of SC.\ Those GICs that exhibit an interlayer band crossing at the $E_F$ turn out to be superconductors with a typical $T_c$ of a few K \cite{Csanyi2005}, and evidence points to such SC being of the conventional phonon-driven type~\cite{Calandra2005Theoretical,Hinks2007Large}.\ The main effect of the partially-filled interlayer band seems to be enhancing the e-ph coupling~\cite{Calandra2005Theoretical,Guzman2014Superconductivity,Yang2024Probing}.

An open question related to the robustness of $d$-wave SC is whether it would be suppressed in real samples of doped graphene by conventional phonon-driven SC \cite{BlackSchaffer2014Chiral}. In analogy to the GICs, phonon-driven SC is also expected to arise in doped single and few-layer graphene if an interlayer band crosses the Fermi level \cite{Profeta2012Phonon,Ludbrook2015Evidence,Chapman2016Superconductivity}, thus possibly overriding the $d$-wave state. 

However, some doped single-layer \cite{Chapman2016Superconductivity,Jugovac2022Clarifying} and few-layer~\cite{Guzman2014Superconductivity, Ichinokura2022Van,Jugovac2022Clarifying} graphene superlattices do not exhibit an interlayer band at the Fermi level, which would suggest a suppression of the phonon-driven SC.\ The absence of the interlayer band at the Fermi level might be explained by its sensitivity to the graphene-dopant distance, which is dependent upon the number of layers and dopant species~\cite{Profeta2012Phonon,Guzman2014Superconductivity}.\ Such sensitivity has been shown most extensively for Li-doped single and few-layer graphene~\cite{Guzman2014Superconductivity,Ichinokura2022Van}. Because in such systems the e-ph coupling is expected to be small, the main effect induced by the ordered dopants is the change in the lattice symmetry of graphene, which significantly alters its electronic spectrum. 

Next, in order to probe the robustness of the $d$-wave state upon changes in the lattice symmetry, we study models for heavily doped graphene superlattices.\ We assume e-ph coupling is not important and focus on Li and Cs doped SLG, which have been realized recently~\cite{Bao2022Coexistence,Ichinokura2022Van,Ehlen2020Origin,Bao2022Coexistence} and exhibit the two most common superlattice symmetries that appear in doped graphene.

\begin{figure*}
\includegraphics[width=1\textwidth]{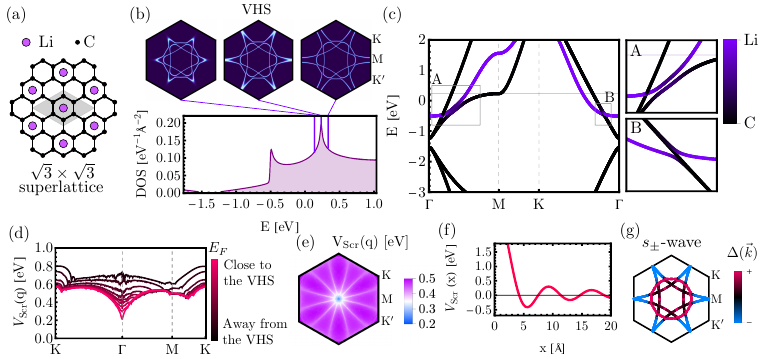}
\caption{
Model for Li-doped SLG and calculated screened electron-electron interaction. (a) $\sqrt{3}\times\sqrt{3}$ graphene-lithium superlattice. (b) DOS and Fermi surfaces at various fillings across the VHS, the latter corresponding to the filling with the highest DOS. High-symmetry points in all panels correspond to the first BZ of the $\sqrt{3}\times\sqrt{3}$ superlattice. (c) Band structure from a tight-binding model considering the $\pi^*$ orbitals of graphene and the $s$ orbital of Li fitted from DFT calculations. The model captures the band folding observed in ARPES measurements \cite{Sugawara2011Fabrication,Toyama2022Two,Bao2022Coexistence}. The A and B insets show the hybridization between carbon and Li bands (indicated by color) at the areas enclosed by the gray rectangles in the main panel.  Experiments in Ref. \cite{Toyama2022Two,Bao2022Coexistence} show fillings close to the horizontal purple line. (d-f) Screened electron-electron interaction. (d) Coulomb interaction along a path in the first BZ. Interaction gets progressively screened as filling gets closer to the VHS. (e) Screened interaction in the full first BZ at the VHS. (f) Screened interaction in real space. (g) Leading OP at $T=10$ mK.}
\label{Fig: LithiumSLG}
\end{figure*}
\subsubsection{Cs-doped SLG}

Recent experiments have shown the presence of a flat band in Cs-SLG~\cite{Ehlen2020Origin}.\ In the experiment, aside from raising $E_F$ close to a VHS, the Cs dopants arrange in a $2\times2$ structure with respect to the original unit cell of SLG, then inducing a BZ folding of the graphene bands.\ We derive a TB model from DFT calculations (see Methods sec.~\ref{sec: DFT} and~\ref{sec: TBFit}) which considers SLG with periodically arranged Cs atoms in a $2\times2$ unit cell, as shown in Fig.~\ref{Fig: CesiumSLG}(a).\ The model considers overlaps between C-C, Cs-Cs and Cs-C, which allow us to account for different hybridization effects.\ Our main results for the electronic band structure are shown in Fig.~\ref{Fig: CesiumSLG}(b-c). Plots show the folded BZ, which corresponds to the $2\times2$ unit cell of the superlattice in Fig. \ref{Fig: CesiumSLG}(a).\ The bands are in good agreement with those obtained from ARPES measurements in Ref.~\cite{Ehlen2020Origin}.\ As shown in Fig.~\ref{Fig: CesiumSLG}(c), Cs $s$-orbitals form a dispersive band that crosses the $E_F$ and hybridizes with the folded $\pi^*$ band of graphene.\ The electronic structure exhibits a VHS at $\Gamma$.\ The maxima in the DOS shown in Fig.~\ref{Fig: CesiumSLG}(b) corresponds to the VHS seen in experiments~\cite{Ehlen2020Origin}.\ We note that this VHS results from the band folding of the original M states of graphene to the $\Gamma$ center of the supercell.\ The momentum-resolved spectral function in the top panel of Fig.~\ref{Fig: CesiumSLG}(b) is highly concentrated at $\Gamma$ when $E_F$ is set at the VHS, while the free-electron like band from the Cs layer causes the larger circular feature centered around $\Gamma$.

The screened Coulomb interaction in real space is shown in Fig.~\ref{Fig: CesiumSLG}(f) with a first minimum at $x \approx 6$~\AA~with an attractive strength of $ V_{Scr}\approx -0.25$~eV which is smaller than that of Tb-SLG. The screened Coulomb interaction for Fermi energies close to the VHS is depicted in Fig.~\ref{Fig: CesiumSLG}(d). While the long-range interaction is increasingly screened as $E_F$ approaches the VHS, we observe that the energy difference between long range ($q\approx\Gamma$) and short range ($q\approx$ M) interactions is significantly smaller than in Tb-SLG. In Cs-SLG the energy difference is $\Delta V_{Scr} \approx 0.2$ eV (in contrast to $\Delta V_{Scr} \approx 1.2$ eV obtained for Tb-SLG), suggesting a minor screening effect. It is noticeable that, even if the Fermi energy is close to the VHS, the variation of the screened potential is small. Because the $T_c$ of the electron-driven SC phase highly correlates with fillings close to a VHS, we estimate the critical temperature at such filling. Despite the high DOS of the original graphene, we find a marginal $T_c$.\ 
An extrapolation of the eigenvalues of the vertex matrix leads to an estimation of a small $T_c\approx 4$ mK.\ We note, that the leading OP at these low temperatures exhibits a non-degenerated $s_\pm$-wave symmetry, as shown in Fig.~\ref{Fig: CesiumSLG}(g).\ These results indicate that the $d$-wave phase might not occur due to the change in lattice symmetry induced by Cs doping, leading to the generation of other OP symmetries.

Although our calculations focused on Cs-doped graphene in order to follow the recent experiments in Ref.~\cite{Ehlen2020Origin}, a similar $2\times 2$ structure might be expected for other dopants, such as K or Rb~\cite{Dresselhaus2002Intercalation}. This has been shown to be the case for graphene bilayers intercalated by K~\cite{Huempfner2023Superconductivity}.\ Additional bands from the extra layers appear at the FS but they are dispersive and are decoupled from each other at the relevant energies, thus we would expect a similar result for those cases. 
It seems that in all of the intercalated few-layer graphene systems that employ these dopants, an interlayer band crosses the Fermi level, suggesting a strong e-ph coupling.  Thus, aside from the $2\times 2$ lattice symmetry being unfavorable for the $d$-wave state, conventional phonon-driven SC is also likely to override it in these systems (this is not always the case for Li intercalated graphene, as we discuss in the next section).\ Therefore, dopants that induce a $2\times2$ superlattice symmetry in graphene, such as Cs, Rb and K, might be detrimental to the $d$-wave superconducting state in graphene.

\subsubsection{Li-doped SLG}
Lithium intercalation has been employed to heavily dope SLG~\cite{Ludbrook2015Evidence,Jugovac2022Clarifying} and also graphene multilayers~\cite{Sugawara2011Fabrication,Toyama2022Two,Bao2022Coexistence,Ichinokura2022Van}.\ Similar to the case of Cs-doping, aside from raising $E_F$, the Li dopants typically arrange in a $\sqrt{3}\times\sqrt{3}$ supercell with respect to the original unit cell of SLG.

Following the same procedure as for Cs-SLG, we derived a TB model that considers the graphene monolayer covered by Li atoms in a $\sqrt{3}\times\sqrt{3}$ periodic arrangement [see Fig.~\ref{Fig: LithiumSLG}(a)].\ The model incorporates C-C, Li-Li, and Li-C hoppings, allowing us to capture the dispersion of the bands and the C-Li hybridization.\ The TB parameters were determined by fitting to the energy dispersion obtained from DFT calculations (see Sections~\ref{sec: DFT} and~\ref{sec: TBFit}). As depicted in Fig.~\ref{Fig: LithiumSLG} (b, c), the dispersion of our model exhibits the expected main features: a $\sqrt{3}\times \sqrt{3}$ band folding and Li-C hybridization, and is in overall good agreement with the dispersion observed in ARPES measurements~\cite{Sugawara2011Fabrication,Ludbrook2015Evidence,Toyama2022Two,Bao2022Coexistence,Ichinokura2022Van}. Plots show the folded BZ, which corresponds to the $\sqrt{3}\times\sqrt{3}$ unit cell of the superlattice in Fig. \ref{Fig: LithiumSLG}(a). The electronic dispersion exhibits a large DOS due to the VHS at the M points, as shown in Fig.~\ref{Fig: LithiumSLG}(b). While the $\sqrt{3}\times\sqrt{3}$ superlattice potential maps the original K and K$^\prime$ points to $\Gamma$, the original M points are mapped to the M points of the folded BZ. The DOS still concentrates at the M points of the folded BZ. Indeed, the momentum-resolved spectral function in the top panel of Fig.~\ref{Fig: LithiumSLG}(b) is highly concentrated at the M points when $E_F$ is set at the VHS (middle panel). However, in contrast to the case of Tb-doped SLG, the VHS in Li-doped graphene doesn't satisfy the condition for an extended VHS \cite{Classen2020Competing}, leading to a singular point, rather than a flat band.

The screened Coulomb interaction in real space is shown in Fig.~\ref{Fig: LithiumSLG}(f) with a first minimum at $x \approx 6$~\AA~with an attractive strength of $ V_{Scr}\approx -0.4$~eV which is smaller than that of Tb-SLG.\ The screened Coulomb interaction for Fermi energies close to the VHS is depicted in Fig.~\ref{Fig: LithiumSLG}(d).\ Results are similar to those of Cs-SLG.\ While the long-range interaction becomes increasingly screened as $E_F$ approaches the VHS, we observe that the energy difference between long range and short range interactions is significantly smaller than in Tb-SLG, although somewhat larger than in Cs-SLG.\ In Li-SLG the energy difference is $\Delta V_{Scr} \approx 0.4$ eV, suggesting a minor screening effect.\ As in Cs-SLG, we also find a marginal $T_c$ close to the VHS.\ An extrapolation of the eigenvalues of the vertex matrix leads to an estimation of a small $T_c\approx 3$~mK.\ At these temperatures, the leading OP exhibits a non-degenerated $s_\pm$-wave symmetry, as shown in Fig.~\ref{Fig: LithiumSLG}(g).\ However, this OP seems to be highly fragile.\ If we slightly shift the $E_F$, the leading OP has a 4-fold degeneracy with symmetries that suggest $p$ or $d$-wave order. Such fragility indicates that different order parameters, strongly dependent on the dopant, can be induced in the $\sqrt{3}\times\sqrt{3}$ superlattice configuration.\ In addition, the dopants that induce a $\sqrt{3}\times\sqrt{3}$ superlattice symmetry in graphene, such as Li, might be detrimental to the $d$-wave state in graphene, since other phases can exist.\ Nevertheless, this might not always be the case for all experimental samples, as reported in Ref.~\cite{Jugovac2022Clarifying}, where high doping by Li intercalation was achieved without an induced superlattice potential nor a Li band crossing $E_F$.\ In that case, the electronic spectrum of Li-SLG resembles that of Tb-SLG, which might lead to a sizable $T_c$. Although for simplicity our calculations focused on Li-SLG, we expect these results will also be applicable to Li-intercalated graphene multilayers, which exhibit the same $\sqrt{3}\times\sqrt{3}$ symmetry~\cite{Bao2022Coexistence,Ichinokura2022Van}.\  Li-doped graphene bilayer might particularly resemble our model, since its $E_F$ is very close to the VHS, and no interlayer band crosses the Fermi level, which could justify neglecting the e-ph coupling \cite{Guzman2014Superconductivity}. The bands introduced by additional layers are unlikely to strongly affect the result, since they are decoupled from the flat band at the relevant energies \cite{Bao2022Coexistence,Ichinokura2022Van}. 
This is in line with recent measurements in Li-intercalated graphene bilayer \cite{Toyama2022Two}, where no SC was found down to 0.8 K.

\section{Discussion}\label{Sec_Discussion}
Several theoretical works have suggested that doped SLG could be a promising platform to realize chiral $d$-wave SC.
A number of methods such as mean-field theory, weak-coupling and functional renormalization group have all led to predictions of the $d$-wave state arising in SLG \cite{Black2007Resonating,Honerkamp2008Density,Gonzalez2008Kohn,Wang2012Functional,Nandkishore2012Chiral,Kiesel2012Competing,BlackSchaffer2014Chiral,Classen2020Competing}.
However, important questions regarding its possible experimental realization have remained an open issue, such as whether the $d$-wave SC survives the strong band renormalizations seen in experiments, its robustness against the source of doping, or whether it will occur at any reasonable $T_c$. Moreover, in part due to uncertainties in model parameters, making quantitative predictions has remained difficult \cite{BlackSchaffer2014Chiral}. 
 
We argue that the $d$-wave SC in SLG is robust against the band renormalizations that occur at dopings close to the VHS, and that it could potentially be realized in heavily doped SLG with a $T_c$ of $\sim 375-600$ mK. Our calculations were performed on a realistic effective model for the electronic structure, based on ARPES measurements on SLG doped beyond the $\pi^*$ VHS via Tb intercalation.\ The theoretical framework we have employed for calculating the $T_c$ and the OP considers SC arising from the strong screening in the electron-electron interaction induced by charge fluctuations.\ This framework has been employed to obtain estimations for the $T_c$ of other graphene superconductors, which have been shown to be in reasonable agreement with experiments across three orders of magnitude. The obtained $T_c$ seems reasonable considering the known critical temperatures for other non-twisted graphene multilayers such as BBG and RTG [see Fig. \ref{Fig: Comparisons}].  

An important issue is whether $d$-wave SC could be overridden by other competing phases, such as conventional phonon-driven SC, or a magnetic state~\cite{BlackSchaffer2014Chiral}.\ Regarding conventional SC, we note that if a doped-SLG system does not exhibit a partially-filled interlayer band, the electron-phonon coupling is expected to be relatively small.\ No interlayer band is found at the $E_F$ for Gd~\cite{Link2019Introducing}, Yb~\cite{Rosenzweig2019Tuning,Rosenzweig2020Overdoping} or Tb doping, and previous experimental works have estimated an electron-phonon coupling constant of $\lambda\approx 0.3-0.4$~\cite{Rosenzweig2020Overdoping,Jugovac2022Clarifying}, which is somewhat smaller than the typical values in the phonon-driven GICs~\cite{Calandra2005Theoretical,Fedorov2014Observation,Ludbrook2015Evidence}. In particular, a value of $\lambda\approx0.3$ is comparable to that of Li-intercalated graphene bilayer~\cite{Guzman2014Superconductivity,Bao2022Coexistence}, which also lacks an interlayer band crossing the $E_F$, and has not been found to be a SC at least above $T_c\approx0.8$ K~\cite{Toyama2022Two}.\ In order to verify the robustness of the $d$-wave state against the competing phonon-driven $s$-wave SC, we have included into our calculations an effective electron-phonon coupling of $\lambda\approx0.49$, 
estimated from our ARPES data (see Methods sec.~\ref{Subsec: phonons} for more details).
We find that the electron-driven $d$-wave state remains unaltered for such $\lambda$. The $d$-wave state is overridden by an $s$-wave state only for values of $\lambda$ beyond a critical value of $\approx0.56$, which is slightly above what we estimate for Tb-doped SLG, and also larger than what has been estimated for Yb~\cite{Rosenzweig2020Overdoping} and Li~\cite{Jugovac2022Clarifying} doping. This critical value is consistent with the experiments in Ref.~\cite{Ludbrook2015Evidence}, where phonon-driven SC was reported to arise in Li-doped SLG, after increasing $\lambda$ to $\approx 0.58$. The competition with other phases is outside of the scope of this work, but has been analyzed in previous works based on renormalization group analysis~\cite{Nandkishore2012Chiral, Kiesel2012Competing, Classen2020Competing}, all pointing out to the $d$-wave state being the leading ground state in some range of experimentally-feasible parameters. Although our calculations have been performed on a model derived from Tb-doped SLG, we expect them to be applicable for other choices of intercalants, such as Gd~\cite{Link2019Introducing}, Yb~\cite{Rosenzweig2020Overdoping}, and others~\cite{McChesney2010Extended,Zaarour2023Flat} that preserve the lattice symmetry of graphene and exhibit a very similar band structure.\

We also performed calculations considering dopants that do change the lattice symmetry of SLG, particularly Li and Cs, and found a drop of at least two orders of magnitude in the $T_c$ and a significant modification in the OP. Variations in the momentum-dependent screened potential, which correlates with KL-type SC, were much less strong in these systems than in Tb-SLG. Moreover, in these systems a large electron-phonon coupling is expected due to a partially-filled interlayer band, and  because additional phonon modes might become accessible to coupling due to the BZ folding, favoring $s$-wave pairing.
These results indicate that employing dopants that change the lattice symmetry of SLG are detrimental to the $d$-wave state.\ The geometry of the FS in SLG, and its relevance for the possible unconventional SC, has been discussed in parallel to the cuprates and the pnictides~\cite{Maiti2013Superconductivity}.

Aside from preserving lattice symmetry, atoms that induce doping to the VHS by pure intercalation lead to better chemical stability, homogeneity, and superior crystallinity~\cite{Link2019Introducing}, compared to those requiring a combination of intercalation and adsorption~\cite{McChesney2010Extended}. This should help to avoid suppression of the $d$-wave state due to disorder \cite{BlackSchaffer2014Chiral,Guo2024Disorder}.\ Thus, Gd~\cite{Link2019Introducing}, Yb~\cite{Rosenzweig2020Overdoping}, Er~\cite{Zaarour2023Flat}, and Tb (this work) seem the most promising dopant choices so far for the realization of $d$-wave SC in SLG.

\section{Methods}

\subsection{Superconductivity from electron-electron interactions} \label{sec: MethodSC}

We consider a diagrammatic technique related to the the KL theory~\cite{Kohn1965New}, where the pairing potential for Cooper pairs is the Coulomb interaction screened by electron-hole excitations.\ In this situation, the SC state emerges from a purely electronic mechanism.\ In contrast to the conventional KL mechanism, however, our approach neglects exchange-like diagrams, but includes all bubble diagrams to infinite orders.\ The RPA is used to calculate the screened Coulomb potential. We follow the procedure described in Ref.~\cite{Cea2022Superconductivity} which we rewrite here in the context of the current work. 

We first consider the Coulomb interaction between two electrons separated by a distance $r$. They experience a $r^{-1}$ long-range repulsion with a potential given by

\begin{equation}
V_{C}(r)=\begin{cases}
\frac{e^{2}}{4\pi\epsilon_{0}\epsilon r} & r\neq0,\\
\frac{\omega_{0}}{\epsilon} & r=0,
\end{cases}
\label{Eq.Vr}
\end{equation}
where $e$ is the electron charge, $\epsilon_0$ the dielectric constant of vacuum and $\epsilon$ the relative dielectric constant of the environment.\ The local repulsion at the same site, denoted as $\omega_0$, is introduced when $r=0$, with an estimated value of around $17$ eV for single-layer graphene~\cite{Wehling2011Strength}.\ We set the dielectric constant $\epsilon=4$ to mimic the screening effect provided by a substrate environment. It is worth noting that $\epsilon$ is the only parameter in the Coulomb potential that is not directly taken from experiment or calculation. However, a prior study~\cite{Li2023Charge} showed a weak dependence between the exact value of $\epsilon$ and the resulting superconducting properties. Since the Coulomb potential changes gradually at the atomic level, we further assume that the interaction between two electrons at different cells only depends on the distance between their centers. Consequently, the Fourier transform of $V_C(r)$ can be expressed as

\begin{equation}
V_C(\vec{q})=\sum_{\vec{R}}V_C(|\vec{R}|)e^{-i \vec{q}\cdot\vec{R}},
\label{Eq.Vc}
\end{equation}
where $r=|\vec{R}|$ is the separation between unit cells and $q$ runs over all points within the BZ. The primitive lattice vectors are $\vec{a}_1$ and $\vec{a}_2$ and $\vec{R} = m \vec{a}_1+n \vec{a}_2$ with $m,n$ integers.\ The number of lattice vectors in the sum of Eq.~\ref{Eq.Vc} is determined by the size of the discrete grid used to sample the BZ.  The form of Eq.~\ref{Eq.Vc} is consistent with the periodicity in momentum space upon translations by reciprocal vectors. The RPA is used to obtain the Coulomb interaction,  

\begin{equation}
    \label{Eq.VScr}
    V_{Scr}(\vec{q})=\frac{V_C(\vec{q})}{1-\Pi(\vec{q})V_C(\vec{q})},
\end{equation}
where $\Pi(q)$ is the static charge susceptibility and is given by

\begin{align}
    \Pi(\vec{q})=\frac{2}{N_k}\sum_{\vec{k}}\sum_{m,n}
    \frac{f(\xi_{n,\vec{k}})-f(\xi_{m,\vec{k}+\vec{q}})}
    {E_{n,\vec{k}}-E_{m,\vec{k}+\vec{q}}} \nonumber \\ \times|\langle u_{n,\vec{k}} | u_{m,\vec{k}+\vec{q}} \rangle|^2,
    \label{Eq.Pi}
\end{align}
with $E_{n,\vec{k}}$ and $|u_{n,\vec{k}}\rangle$ being the energy and eigenvector corresponding to the $n$-th band at wave-vector $\vec{k}$. $f(\xi_{n,\vec{k}})=[1+\exp(\xi_{n,\vec{k}}/k_B T)]^{-1}$ is the Fermi-Dirac distribution at temperature $T$ and chemical potential $\mu$ with $\xi_{n,\vec{k}}=E_{n,\vec{k}}-\mu$, and $N_k$ is the number of cells or number of $\vec{k}$ points used to sample the BZ.

Following~\cite{Cea2022Superconductivity}, one can derive a linearized gap equation for the screened Coulomb interaction of the form
\begin{equation}
    \Delta_{m_1m_2}(\vec{k})=\sum_{\vec{k}'n_1n_2}\Gamma_{m_1m_2,n_1n_2}(\vec{k},\vec{k}')\Delta_{n_1n_2}(\vec{k}'),
    \label{Eq.Delta}
\end{equation}
where the superconducting critical temperature and order parameter $\Delta(\vec{k})$ can then be obtained by diagonalizing the hermitian kernel,

\begin{multline}
 \Gamma_{m_1 m_2, n_1 n_2} (\vec{k},\vec{k}')= \\
-\frac{1}{N_k}V_{Scr}(\vec{k}-\vec{k}') \langle u_{m_1, \vec{k}}|u_{n_1, \vec{k}'}\rangle
        \langle u_{n_2, \vec{k}'}|u_{m_2, \vec{k}}\rangle \\
 \times\sqrt{\frac{f(-\xi_{m_2, \vec{k}})-f(\xi_{m_1,\vec{k}})}{\xi_{m_2,\vec{k}}+\xi_{m_1,\vec{k}}}}
\sqrt{\frac{f(-\xi_{n_2, \vec{k}'})-f(\xi_{n_1,\vec{k}'})}{\xi_{n_2,\vec{k}'}+\xi_{n_1,\vec{k}'}}}.
\label{Eq.Gap}
\end{multline}
The superconducting $T_c$ corresponds to the highest temperature at which the largest eigenvalue of the kernel $\Gamma$ equals one, and the OP is the corresponding eigenvector. 

This framework has produced estimations of the $T_c$ for the graphene superconductors that are in good agreement with experiments across three orders of magnitude, as shown in Fig.~\ref{Fig: Comparisons}. The error bars indicate the variation in the calculated $T_c$ obtained within different expressions for the Coulomb interaction or by employing continuum or TB model hamiltonians~\cite{Cea2022Superconductivity,Cea2023Superconductivity,Jimeno2023Superconductivity,Li2023Charge}. All calculations of $T_c$ are done at VHS filling. For non-twisted graphene systems, the consideration of electron-phonon coupling has been shown to not significantly change the resulting SC \cite{Li2023Charge} while for the twisted systems, moir\'e-induced Umklapp scatterings give an important contribution and have to be included \cite{Cea2021Coulomb,Long2024Evolution}.
  
\subsection{Density Functional Theory calculations} \label{sec: DFT}
First-principles calculations were carried out using a numerical atomic orbitals approach to density functional theory (DFT)~\cite{kohsha1965,HohKoh1964}, which was developed for efficient calculations in large systems and implemented in the \textsc{Siesta} code~\cite{ArtAng2008,SolArt2002,siesta-2020}. 
We have used the generalized gradient approximation (GGA) and, in particular, the functional of Perdew \emph{et al.}~\cite{PBE96}.
Only the valence electrons are considered in the calculation, with the core being replaced by norm-conserving scalar relativistic pseudopotentials~\cite{vansetten2018dojo,garcia2018psml}.
We use the Grimme semiempirical method to correctly describe the distance between the graphene layer and the alkali atom~\cite{grimme2006semiempirical}.
The non-linear core-valence exchange-correlation scheme~\cite{LFC82} was used for all elements. 
We have used a split-valence double-$\zeta $ basis set including polarization functions~\cite{arsan99} for C and Cs and a split-valence double-$\zeta $ basis set for Li. 
The energy cutoff of the real space integration mesh was set to 1000 Ry. 
To build the charge density (and, from this, obtain the DFT total energy and atomic forces), the Brillouin zone (BZ) was sampled with the Monkhorst-Pack scheme~\cite{MonPac76} using grids of (11$\times$11$\times$1).
The crystal structure of graphene was fully optimized, obtaining a lattice constant for the hexagonal lattice of 2.49 \AA. 
Then, we built the two superstructures, $2\times2$ and $\mathrm{R}30^\circ \sqrt{3} \times \mathrm{R}30^\circ \sqrt{3}$ for Cs and Li, respectively.
We relaxed the structure, only allowing for the alkali atoms to move in the $z$-direction with a threshold of 0.01 eV/\AA.

\subsection{Tight-binding models} \label{sec: TBFit}

\begin{figure*}
\includegraphics[width=\textwidth]{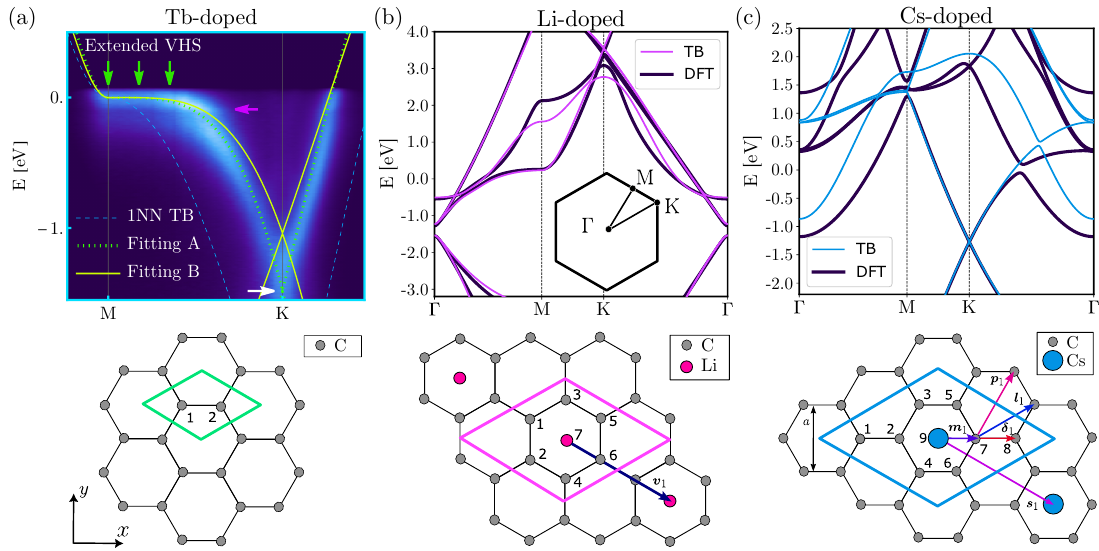}
\caption{(a) Top: Electronic bands of the TB models for Tb-doped SLG fitted from ARPES measurements. Bottom: Atomic basis for the TB model. ARPES measurements show a kink (purple arrow) below $E_F$ that has been ascribed to mass renormalization due to the electron-phonon interaction (see Methods sec. \ref{Subsec: phonons}). (b,c) Top: electronic band structures comparing DFT calculations and fitted TB bands for Li doped graphene (b) and Cs doped graphene (c). Inset in (b) shows the path followed in the first BZ. High-symmetry points at each panel correspond to the BZ of the $1\times1$, $\sqrt{3}\times\sqrt{3}$ and $2\times2$ lattices, respectively. Bottom: Atomic basis and relevant neighbour vectors for the TB models. In pink and blue, supercells formed by $\textbf{a}_1$ and $\textbf{a}_2$ in each geometry.}
\label{Fig: TB}
\end{figure*}

As mentioned in sec.~\ref{Sec: results-tbslg}, the renormalization of the bands due to high doping can be attributed to a combination of many-body interaction phenomena. However, we can effectively take it into account by fitting an effective single-particle TB Hamiltonian for electrons in a graphene lattice with up to 3NN hopping~\cite{McChesney2010Extended,Classen2020Competing}. 
In this model, the requirement of an extended VHS fixes the 3NN hopping $t_3$ as a function of the first- ($t$) and second-nearest neighbor hopping ($t_2$) so that the number of free parameters reduces to three ($t$, $t_2$, and chemical potential $\mu$).\ Hence, the model is well defined just via three ARPES band-structure hallmarks, i.e., the binding energies of the Dirac point ($1.55$ eV) and the VHS ($0.07$ eV) as well as the effective mass of the $\pi^*$ band along $\mathrm{M}\Gamma$ ($0.15$ $m_e$).
    The electronic bands of the TB models for two different fittings (parameters are reported in Table~\ref{tab:tableTerbium}) of the Tb-doped SLG are shown in Fig.~\ref{Fig: TB}(a). 
Both dispersions satisfy the requirement for an extended VHS~\cite{Classen2020Competing} featured in the ARPES measurements. 
Fitting A (also shown in Fig. \ref{Fig: TerbiumSLG}) is able to reproduce the full band from the VHS down to the Dirac point (white arrow), but misses the additional flatness in the band that arises from the e-ph kink (purple arrow). 
On the other hand, fitting B leads to a band that fits the flatness due to the kink to a larger extent, but significantly deviates from the ARPES measurements for lower energies. Superconductivity was computed for both fittings, A and B. Both fittings lead to the same OP, with the $d$-wave orders essentially doubly-degenerate. Close to $T_c$, the $d_{xy}$ and $d_{x^2-y^2}$ orders have approximately equal eigenvalues, with the former being slightly larger ($\approx$1.05 vs $\approx$1.00). However, these two eigenvalues are quite separated from the next largest eigenvalue ($\approx 0.6$), which has a different symmetry. 
The $T_c$ calculated with fitting A is $T_c\approx370$ mK, and with fitting B it increases by a factor of $\approx 1.8$ ($T_c\approx600$ mK), indicating the robustness of the $d$-wave state.

For the cases of Li and Cs doping, atomistic calculations were carried out using an effective TB model fitted from the DFT calculations. 
Previous works~\cite{Ehlen2020Origin}, have only included nearest neighbours carbon-carbon hoppings in their TB, which results in an oversimplified model of the system which is not able to capture key features of the DFT and experimental band structures.
This fact hinders those models from obtaining a good prediction for the critical temperature of doped graphene.
Here, we employ a more realistic model that accounts for higher-order neighbours which results in a better description of the band structure of the systems.
In the following, we describe our model:
The basis set for our Hamiltonian is composed of a $p_z$ orbital for the carbon atoms and an $s$ orbital for the Cs/Li atom.
Since the Cs atoms arrange in a $2\times2$ supercell, we have a 9-orbital basis.
On the other hand, for the Li case they arrange in a ($\sqrt{3}\times\sqrt{3}$)R30$^{\circ}$ supercell, and we have a 7-orbital basis.
We define the lattice vectors for the Cs case, $\{\boldsymbol{a}_i^s \}$, and for the Li case $\{\boldsymbol{a}_i^l \}$ as:
\begin{align}
\textbf{a}_1^s & = a(\sqrt{3},1)\\
\textbf{a}_2^s & = a(\sqrt{3},-1)\\
\textbf{a}_1^l & = a(\sqrt{3}/2,3/2)\\
\textbf{a}_2^l & = a(\sqrt{3}/2,-3/2),
\end{align}
where $a=2.46$ \AA~is the graphene lattice constant.
In addition, we define the vectors of the C-C nearest neighbours $\{\boldsymbol{\delta}_i \}$ with hopping amplitude $t$, C-C second nearest neighbours $\{\boldsymbol{l}_i \}$ with hopping amplitude $t_2$, third C-C nearest neighbours $\{\boldsymbol{p}_i \}$ with hopping amplitude $t_3$, Cs-Cs nearest neighbours $\{\boldsymbol{s}_i \}$ with hopping amplitude $t_s$, Li-Li nearest neighbours $\{\boldsymbol{v}_i \}$ with hopping amplitude $t_l$ and C-Cs/Li nearest neighbours $\{\boldsymbol{m}_i \}$ with hopping amplitude $t_s'$ for the Cs and $t_l'$ for the Li. Additionally, we found that a second nearest neighbour interaction between Li atoms improved considerably the fitting of the Li-band and thus a $\{2\boldsymbol{v}_i \}$ with hopping amplitude $t_{2l}$ was also used:
\begin{align}
\boldsymbol{\delta}_1 & = a(1/\sqrt{3},0), \text{ $\boldsymbol{\delta}_{i+1}=\hat{C_3}\boldsymbol{\delta}_{i}$}\\
\boldsymbol{l}_1 & = \boldsymbol{\delta}_1-\boldsymbol{\delta}_3, \text{ $\boldsymbol{l}_{i+1}=\hat{C_6}\boldsymbol{l}_{i}$}\\
\boldsymbol{p}_1 & = \boldsymbol{l}_1+\boldsymbol{\delta}_2, \text{ $\boldsymbol{p}_{i+1}=\hat{C_3}\boldsymbol{p}_{i}$}\\
\boldsymbol{m}_1 & = \boldsymbol{\delta}_1, \text{ $\boldsymbol{m}_{i+1}=\hat{C_6}\boldsymbol{m}_{i}$}\\
\boldsymbol{s}_1 & = \textbf{a}_1^s, \text{ $\boldsymbol{s}_{i+1}=\hat{C_6}\boldsymbol{s}_{i}$}\\
\boldsymbol{v}_1 & = \textbf{a}_1^l+\textbf{a}_2^l, \text{ $\boldsymbol{v}_{i+1}=\hat{C_6}\boldsymbol{v}_{i}$}
\end{align}
where the operator $\hat{C_n}$ is an anticlockwise rotation of $\theta=2\pi/n$:
\begin{equation}
\hat{C_n}=\begin{pmatrix}
\text{cos}(\theta) & -\text{sin}(\theta) \\
\text{sin}(\theta) & \text{cos}(\theta) 
\end{pmatrix}
\end{equation}\\

The TB Hamiltonian, 
\begin{equation}
    \hat{H}_0  =\sum_{i,j,\textbf{k}}h_{ij}(\textbf{k})c_{i\textbf{k}}^\dagger c_{j\textbf{k}},
\end{equation}
where $c^\dagger_{i\textbf{k}}$ creates an electron in the orbital $i$ and $c_{j\textbf{k}}$ annihilates one in the orbital $j$, can be decomposed in the diagonal and off-diagonal terms.
For the Cs case, the diagonal elements can be written as:
\begin{align}
    h_{ii}(\textbf{k}) & =\begin{cases}    
        \epsilon_{c} & i = 1,8 \hspace{1mm}\\
        \epsilon_{c2} & i=2,3,4,5,6,7\hspace{1mm}\\
        \epsilon_s+t_s\sum_{n=1}^6e^{i\textbf{k}\textbf{s}_n} & i = 9  \hspace{4mm}\\
    \end{cases}
\end{align}
where $\epsilon_i$ are the on-site energy for the C and Cs atoms. 
We distinguish between the on-site energy of C atoms surrounding the Cs atom ($\epsilon_{c2}$) and those without the influence of the heavy atom ($\epsilon_{c}$). 
This distinction arises from the physical intuition that the C atoms surrounding the Cs atom suffer a different effective potential than those that are further away from the Cs atom.

For the Li case, these terms can be written as:
\begin{align}
    h_{ii}(\textbf{k}) & =\begin{cases}   
    \epsilon_{l}       & i = 1,4,5 \\
    \epsilon_{l2}      & i = 2,3,6   \\
    \epsilon_l+t_l\sum_{n=1}^6e^{i\textbf{k}\textbf{v}_n} &  i = 7  
    \end{cases}
\end{align}
where $\epsilon_l$ are the on-site energy for the C and Li atoms.
For the Li-case we find that, in order to open the gap in the TB observed at $\Gamma$ in the DFT calculations, we need to break $C_6$ symmetry by imposing a different on-site energy between the 
C atoms. 
Thus atoms 1, 4 and 5  have an onsite term ($\epsilon_{l}$) and 2, 3 and 6, ($\epsilon_{l2}$). 

The off-diagonal terms for the Cs and Li cases can be written as:
\begin{align}
    h_{ij}&(\textbf{k})  = t\sum_{n=1}^3f(\textbf{r}_{ij}-\boldsymbol{\delta}_n)e^{i\textbf{k}\boldsymbol{\delta}_n}+t_2\sum_{n=1}^6f(\textbf{r}_{ij}-\boldsymbol{l}_n)e^{i\textbf{k}\boldsymbol{l}_n}\nonumber\\
    & +t_3\sum_{n=1}^6f(\textbf{r}_{ij}-\boldsymbol{p}_n)e^{i\textbf{k}\boldsymbol{p}_n}+t_{s/l}'\sum_{n=1}^6f(\textbf{r}_{ij}-\boldsymbol{m}_n)e^{i\textbf{k}\boldsymbol{m}_n},
\end{align}
where $\textbf{r}_{ij}$ is the interatomic distance connecting sites $i$ and $j$ and
\begin{equation}
    f(x)=
    \begin{cases}
        1 & x = 0\\
        0 & x \neq 0 
    \end{cases}
\end{equation}

\begin{figure*}
\includegraphics[width=1.0\textwidth]{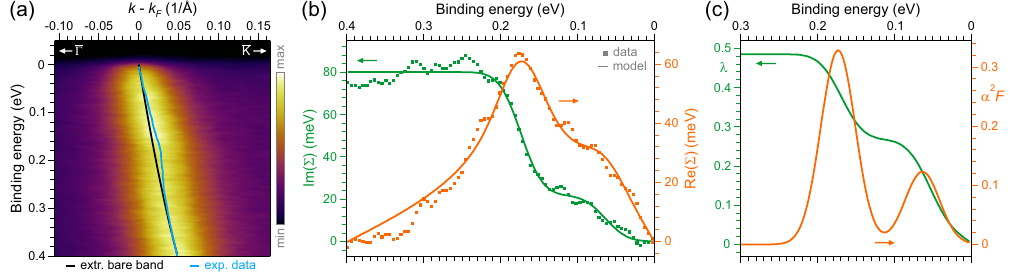}
\caption{Electron-phonon coupling of Tb-intercalated SLG. (a) Close-up of the low-energy $\pi^*$ dispersion along $\overline{\Gamma\mathrm{K}}$ (light blue curve) revealing a prominent kink $\approx0.17$ eV below $E_F$. The bare band extracted via the algorithm of Ref.\ \cite{Pletikosi2012Finding} is also overlaid (black curve). (b) Corresponding real (Re, orange markers, right axis) and imaginary (Im, green markers, left axis) parts of the spectral function $\Sigma$. Im($\Sigma$) is shifted by a constant offset of $0.29$ eV. Im($\Sigma$) is modelled by two step-like increases, corresponding to two phonon modes with energies of $64$ and $173$ meV (green curve). Its Kramers-Kronig transform (orange curve) matches Re($\Sigma$), hence demonstrating self-consistency. (c) From the modelled Im($\Sigma$), we determine the Eliashberg function $\alpha^2 F(E)=\frac{1}{\pi}\frac{\mathrm{d}}{\mathrm{d}E}\mathrm{Im}(\Sigma)$ (orange curve, right axis) and from it the electron-phonon coupling constant $\lambda(E)=2\int\mathrm{d}E\,(\alpha^2F/E)$ (green curve, left axis) with a maximum value of $\lambda=0.49$.}
\label{Fig: elph}
\end{figure*}

A gradient descent approach was used to fit the band structure obtained with DFT and extract the amplitude of the different hoppings of the model. We define the TB eigenenergies as a function of the different on-site energies and interatomic hoppings. Thus, the TB energy for a given band $n$ at a point $\textbf{k}$ depends on the on-sites $\epsilon$ and hopping parameters $t$:
\begin{equation}
    \varepsilon_{n,\textbf{k}} = \varepsilon_{n,\textbf{k}}\left(\epsilon, t \right)
\end{equation}
We now can define a cost error function as the difference between the TB and DFT energies as:

\begin{equation}
\begin{split}
    C\left( \epsilon, t \right) & = \frac{1}{N_{b} N_{\textbf{k}}}\sum_{n=1}^{N_b}{\sum_{\textbf{k}}}[E^{DFT}_{n, \textbf{k}} - \varepsilon_{n, \textbf{k}}\left(\epsilon, t \right)]^{2},
\end{split}
\end{equation}
where $N_b$ is the number of bands used for the fitting and $N_k$ the number of k-points used in the path $\Gamma-M-K-\Gamma$ [see inset of Fig. \ref{Fig: TB} (b)]. 
In our calculations we used $N_b=4$ bands around the Fermi level. 
The best set of parameters are reported in Table \ref{tab:table2} and the corresponding band structures are shown in Figs. \ref{Fig: TB}(b) and (c) for Li and Cs, respectively.
We can see that our TB model gives accurate results and, most importantly, captures the important features from the DFT band structure such as the dispersion of the C bands [see Fig.~\ref{Fig: CesiumSLG}(c) and ~\ref{Fig: LithiumSLG}(c)] for the orbital contribution]. 
Furthermore, in the Li case the TB model is able to describe the breaking of degeneracy of the bands in the $\Gamma-$M path.
Importantly, it also has a very good agreement with the DFT band structure at the Fermi level since it captures all the degeneracies and dispersions.
All in all, this model and its parameters are a good starting point for calculations that go beyond the single-particle approach.

\begin{table}
\caption{\label{tab:tableTerbium}
Fitting parameters for the 3NN TB models of Tb-doped SLG. Fitting A corresponds to the model shown in Fig. \ref{Fig: TerbiumSLG}. Fittings are compared in Fig. \ref{Fig: TB} (a).}
\begin{ruledtabular}
\begin{tabular}{cccccccccc}
 Model & $t$ [eV] & $t_2$ [eV]& $t_3$ [eV]\\
\hline
Fitting A& -4.077 & -0.925 & -0.557 \\
Fitting B& -3.02 & -0.6 & -0.46\\
\end{tabular}
\end{ruledtabular}
\end{table}
\begin{table}
\caption{\label{tab:table2}
Table with all fitted TB parameters. All values are in units of $t_0=-2.7$ eV which is the graphene first neighbours typical hopping value. Fittings are shown in Fig. \ref{Fig: TB} (b) and (c).}
\begin{ruledtabular}
\begin{tabular}{cccccccccc}
 Atom & $\epsilon_c$ & $\epsilon_{c2}$ & $\epsilon_{s}$ & $t$ & $t_2$ & $t_3$ & $t_{s/l}$ &$ t_{2\;s/l}$ & $t'$\\
\hline
Cs& 0.5 & 0.52 & -0.4 & 1.1 & 0.015 & 0.1 & 0.12 & 0 & 0.025\\
Li& 0.695 & 0.81 & -0.38 & 1.28 & 0.08 & 0.2 & 0.135 & -0.04 & 0.05\\
\end{tabular}
\end{ruledtabular}
\label{table2}
\end{table}

\subsection{Preparation of highly-doped, quasi-freestanding SLG via Tb intercalation} \label{sec: graphene-exp}
On-axis, single crystalline, $n$-doped 6H-SiC(0001) wafer segments (SiCrystal GmbH) were used as substrates for graphene growth. The substrates were first etched with molecular hydrogen at $1400$~$^{\circ}\mathrm{C}$ and near ambient pressure to obtain atomically flat terraces~\cite{ramachandran1998,soubatch2005}. Following the well-established method of Emtsev \emph{et al.} \cite{emtsev2009}, graphene was grown by heating the substrates to around 1450~$^{\circ}\mathrm{C}$ for about $5$ min in $800$ mbar argon atmosphere. This yields several-$\mu$m-wide terraces, uniformly covered with the $(6\sqrt{3}\times6\sqrt{3})\mathrm{R}30^\circ$ carbon buffer layer reconstruction. This so-called zerolayer graphene (ZLG) does not yet present the electronic properties of freestanding single-layer graphene \cite{emtsev2008,riedl2010} due to covalent bonding to the Si-terminated SiC substrate. Both, hydrogen etching and surface graphenization were performed \emph{ex situ} in an inductively heated reactor hosting a graphite susceptor.

The as-grown ZLG/SiC samples were transferred into ultrahigh vacuum (UHV) and first degassed at $700$~$^{\circ}\mathrm{C}$ for about $20$ min. Tb was evaporated from a commercial Knudsen cell (OmniVac) onto ZLG at a rate of about 1.7~\AA{}/min. A first cycle of Tb deposition was performed for $8$ min with the sample at room temperature. Subsequently, the samples underwent $10$--$20$ min long sequential annealing cycles from $400$--$900$~$^{\circ}\mathrm{C}$ in steps of $100$~$^{\circ}\mathrm{C}$. This was followed by $<10$ s of flash-annealing to $1100$--$1150$~$^{\circ}\mathrm{C}$. High temperatures and short times are necessary in order to desorb excess Tb left on top of ZLG while avoiding the growth of additional graphene layers in UHV. At this stage, partial intercalation is achieved as observed by low-energy electron diffraction (LEED). The LEED pattern contains a mixture of ZLG and SLG features indicating that in some parts of the surface Tb atoms have migrated to the ZLG/SiC interface and saturate the substrate dangling bonds so that patches of quasi-freestanding SLG are formed. The intercalation was finalized by another deposition cycle of Tb, this time for $\approx 10$ min at an elevated sample temperature of $600$~$^{\circ}\mathrm{C}$, followed by the very same annealing steps as in the first cycle. This two-step preparation process reliably resulted in homogeneously intercalated SLG.

\subsection{ARPES measurements} \label{sec: arpes}
Synchrotron-based ARPES was carried out at the BLOCH beamline of MAX IV Laboratory in Lund, Sweden. The endstation hosts a DA30-L hemispherical analyzer (Scienta Omicron), capable of recording 2D photoelectron intensity distribution maps via electronic deflection perpendicular to the entrance slit. The latter is oriented perpendicular to the plane of light incidence. Our Tb-intercalated samples have been kept under true UHV conditions during the transfer by means of a dedicated vacuum transport suitcase (Ferrovac).

The dataset of Fig.~\ref{Fig: TerbiumSLG}(b) and (c) was obtained for a sample temperature of $\approx20$ K, using linear horizontally polarized light with a photon energy of $65$ eV and a beam spot size of about $14\times 7$ $\mu\mathrm{m}^2$. The angular resolution was better than $0.3^\circ$ and the combined total energy resolution (analyzer and beamline) was set to $\approx15$ meV.

The recorded photoelectron intensity distribution map covers an area of about $2.5\times1.5$ \AA{}$^{-2}$ centered on the $\mathrm{KMK'}$ border of the first BZ. The map was then mirror-symmetrized with respect to (i) the $\mathrm{KMK'}$ line---determined with very high precision---and (ii) the perpendicular axis through $\mathrm{M}$. To a large extent, this removes the so-called dark corridor of the $\pi^*$ ($\pi$) band whose intensity is otherwise suppressed inside the first (repeated) BZ due to matrix-element effects \cite{Gierz2011Dark}. A sector with an opening angle of $60^\circ$ (spanned by $\mathrm{K}\Gamma\mathrm{K'}$) is then sequentially repeated to visualize the Fermi surface in Fig.\ \ref{Fig: TerbiumSLG}(c).

In the low-energy $\pi^*$ dispersion along the $\overline{\Gamma\mathrm{K}}$ direction, cf.\ raw data in Fig.~\ref{Fig: elph}(a), renormalization effects by electron-phonon coupling can be observed, manifested by the well-known electron-phonon kink. Using an algorithm reported in Ref.\ \cite{Pletikosi2012Finding}, the bare band can be extracted from the band position and band width (FWHM). Consistency of the procedure is monitored by Kramers-Kronig transformation between the corresponding real and imaginary parts of the spectral function $\Sigma$, see Fig.~\ref{Fig: elph}(b). From the modelled Im($\Sigma$), the Eliashberg function $\alpha^2 F(E)=\frac{1}{\pi}\frac{\mathrm{d}}{\mathrm{d}E}\mathrm{Im}(\Sigma)$ and subsequently, the electron-phonon coupling constant $\lambda(E)=2\int\mathrm{d}E\,(\alpha^2F/E)$ are determined. The result is demonstrated in Fig.~\ref{Fig: elph}(c). $\lambda$ reaches a maximum value of $0.49$.

\subsection{Electron-phonon interactions}\label{Subsec: phonons}
 An open problem regarding the feasibility of the $d$-wave state in graphene is whether electron-phonon interactions are strong enough to override the electron-driven $d$-wave state in favor of a conventional phonon-driven $s$-wave state~\cite{BlackSchaffer2014Chiral}. In order to test the robustness of the $d$-wave state, we have included an attractive electron-phonon interaction $g_{e-ph}$ into our calculations. The $d$-wave state remains unaltered as long as $|g_{e-ph}|$ stays below a critical value of $\approx 1.3$ eV. Beyond this value, the order parameter changes from $d$-wave to $s$-wave,
as expected for an attractive phonon-driven interaction. We have estimated the expected value of $|g_{e-ph}|$ for the Tb-doped SLG and find it to be about $\approx1.1$ eV, indicating that the electron-phonon interactions do not override the $d$-wave state.

We also have considered ARPES measurements of the electronic dispersion of heavily-doped graphene, which consistently reveal a kink around $0.1-0.2$ eV below $E_F$, as shown in Fig.~\ref{Fig: elph}. This kink has been ascribed to mass renormalization due to the electron-phonon interaction, with an estimated coupling of $\lambda\approx0.3-0.4$ for other dopants~\cite{Park2008Van,McChesney2010Extended,Rosenzweig2020Overdoping,Jugovac2022Clarifying}. From our ARPES measurements on Tb-doped SLG, we estimate $\lambda\approx 0.49$ (see Methods sec.~\ref{sec: arpes} and Fig.~\ref{Fig: elph}). In order to estimate $g_{e-ph}$, we replace $V_{Scr}$ by $g_{e-ph}$ in the kernel of Eq. \ref{Eq.Gap} and look for the $g_{e-ph}$ value that produces the ARPES-derived $\lambda$. For a given value of $g_{e-ph}$, $\lambda$ is obtained by fitting the largest eigenvalue of the kernel as function of temperature (for $T>T_c$) to a curve of the form $\lambda \log (W/k_B T)$. With this procedure we find $g_{e-ph}\approx -1.1$ eV. Such interaction leads to $\lambda\approx 0.49$ and $W\approx 15$ meV. The obtained value of $W$ may be understood as the width of the flat band, as it coincides with the energy cutoff around $E_F$ for which Eq.~\ref{Eq.Gap} converges. 
For other dopants with $\lambda\approx 0.3-0.4$~\cite{Park2008Van,McChesney2010Extended,Rosenzweig2020Overdoping,Jugovac2022Clarifying}, we estimate a smaller value of $|g_{e-ph}|\approx 0.4-0.8$ eV.

Alternatively, we can use a microscopic calculation, where we define the coupling of the optical phonons at $\Gamma$ to the electrons through the modulation of the nearest neighbor hopping, $t$, by the changes in bond lengths induced by the phonon displacements. There are two degenerate phonons at $\Gamma$ \cite{Ferrari2006Raman}. The displacements are of opposite signs in the two atoms of the unit cell, and the phonons are polarized along the $x$ and $y$ axes~\cite{Bostwick2007Quasiparticle,Tse2007Phonon,Forti20011Large,Pletikosi2012Finding}. The dependence of $t$ on bond length is characterized by a dimensionless quantity:
\begin{align}
    \beta &= \frac{a}{t} \frac{\partial t}{\partial a} \approx 3
    \label{beta}
\end{align}
where $a$ is the bond length. Phonons at $\Gamma$ induce an attraction within the three nonequivalent M points in the Brillouin zone where the van Hove singularities reside. The mean square displacement of a given atom is:
\begin{align}
    \langle | \Delta \vec{r} |^2 \rangle &= \frac{\hbar}{2 M_C \omega_\Gamma}
    \label{ZPM}
\end{align}
where $M_C$ is the mass of the Carbon atom. These displacements induce changes in the three inequivalent bonds of the honeycomb lattice equal to $\{ 2 | \Delta \vec{r} | , - | \Delta \vec{r} |   , - | \Delta \vec{r} |  \}$. The associated changes in the hoppings lead to changes in the band energy at the three inequivalent M points equal to 
\begin{align}
  \Delta \epsilon_M &= \beta \times \frac{t}{a}  \times \{  4 | \Delta \vec{r} | , - 2 | \Delta \vec{r} |   , - 2 | \Delta \vec{r} |  \}
  \label{eph_coupling}
\end{align}
which leads to an average electron-phonon coupling:
\begin{align}
    g_{e-ph} &= -2 \times \left\langle \frac{\Delta \epsilon_M^2}{\hbar \omega_\Gamma} \right\rangle_M = -\frac{8 \beta^2 t^2}{M a^2 \omega_\Gamma^2}.
    \label{geph}
\end{align}
We take $\hbar \omega_\Gamma \approx 0.17 $ eV. The value of $t$ for undoped graphene is $ \approx 2.7$ eV. The heavily doped graphene studied here is described by a strongly renormalized
$t$, from $\approx1.5$ eV (if fitting to 1NN) to $\approx4.077$ eV, see Table  [\ref{tab:tableTerbium}].
As a result, the value of $g_{e-ph}$ varies over a wide range:

\begin{align}
    g_{e-ph} ( t = 4\,{\rm eV} ) &\approx- 6.8 \, {\rm eV} \nonumber \\
    g_{e-ph} ( t = 1.5 \, {\rm eV} ) &\approx -0.96 \, {\rm eV}
\end{align}

This estimation range includes the value extracted from ARPES experiments mentioned before. As optical modes are not screened, we add the bare electron-phonon interaction  to the electron-electron contribution in Eq.~\ref{Eq.Gap}, so the final interaction is $V_{Scr}(q)+g_{e-ph}$. 

Considering only the $g_{e-ph}$ coupling in Eq.~\ref{Eq.Gap}, the resulting OP is always $s$-wave, as expected for a constant, attractive interaction. However, when including also the electron-electron contribution $V_{Scr}(q)$, the resulting OP is $d$-wave (as shown in Fig.~\ref{Fig: TerbiumSLG}) as long as $|g_{e-ph}|$ stays below the critical value of $\approx1.3$ eV (which gives $\lambda\approx0.56$). This is consistent with the experiments in Ref.~\cite{Ludbrook2015Evidence}, where phonon-driven SC was reported to arise in Li-doped SLG after increasing $\lambda$ to $\approx 0.58$. The critical value of $|g_{e-ph}|$ coincides with the value of the screened interaction at the VHS, $V_{Scr}(\vec{q}=\text{M})\approx1.3$ eV [see Fig.~\ref{Fig: TerbiumSLG}(d)]. Thus, if $|g_{e-ph}|>V_{Scr}(q=\text{M})$, the OP turns into $s$-wave, with the same sign at the three M-points.

For other dopands, a lower $\lambda\approx0.3-0.4$~\cite{Park2008Van,McChesney2010Extended,Rosenzweig2020Overdoping,Jugovac2022Clarifying} has been estimated. It should be noted, that for Gd, the flat band scenario develops without a superperiodicity. However, by further annealing, a weak $\sqrt{3}\times\sqrt{3}$ contribution can be observed with eVHS filling maintained. When analyzing the band width in this later state using the same procedure as for Tb, the low energy phonons contribute a $\lambda$ of $\approx$ 0.36, while the high energy (170 meV) regime indeed only contributes 0.19. So, in total this amounts to $\lambda$ $\approx$ 0.54, which is still below the calculated limit~\cite{LinkThesis}. Yet, as noted, the flat band situation develops without a supercell. In fact, also for the Tb case, a considerable coupling is observed for the low energy phonon regime even without a $(\sqrt{3}\times\sqrt{3})$ superstructure. We speculate that it may actually be caused by residuals from the intrinsic $(6\sqrt{3}\times6\sqrt{3})$ superstructure of epitaxial graphene on SiC(0001) which is not visible in LEED after the intercalation but -- of course -- naturally still present.


These estimations presented above and supported by our ARPES results, indicate that the electron-phonon coupling in VHS-doped graphene is not strong enough to override the $d$-wave SC.

\textit{Acknowledgments}. We thank Zhen Zhan, Alejandro Jimeno-Pozo, Hector Sainz-Cruz, Min Long, Adri\'an Ceferino and Debmalya Chakraborty for fruitful discussions.\ IMDEA Nanociencia acknowledges support from the ‘Severo Ochoa’ Programme for Centres of Excellence in R\&D (CEX2020-001039-S/AEI/10.13039/501100011033). 
G.P.-M., J.A.S.-G., P.A.P. and F.G. acknowledge support from NOVMOMAT, project PID2022-142162NB-I00 funded by MICIU/AEI/10.13039/501100011033 and by FEDER, UE as well as financial support through the (MAD2D-CM)-MRR MATERIALES AVANZADOS-IMDEA-NC.
G.P.-M. is supported by Comunidad de Madrid through the PIPF2022 programme (grant number PIPF-2022TEC-26326).
P.R., B.M., K.K., and U.S.\ acknowledge support by the Deutsche Forschungsgemeinschaft (DFG, German Research Foundation) through Projects Ku 4228/1-1 and Sta 315/13-1 within the Research Unit FOR5242. MAX IV Laboratory is acknowledged for time on beamline BLOCH under proposal 20221217. Research conducted at MAX IV, a Swedish national user facility, is supported by the Swedish Research Council under contract 2018-07152, the Swedish Governmental Agency for Innovation Systems under contract 2018-04969, and Formas under contract 2019-02496. S.A.H. and G.G.N. acknowledge financial support from UNAM DGAPA PAPIIT IN101924 and CONAHCyT project 1564464.

\bibliographystyle{apsrev4-2}
%


\end{document}